\documentclass[prd,twocolumn,twoside,preprintnumbers,superscriptaddress]{revtex4}

\usepackage{amsmath}
\usepackage{graphicx}
\usepackage{dcolumn}
\usepackage[hyperfootnotes=false]{hyperref}
\usepackage{xspace}
\usepackage{color}
\usepackage{url}

\newcommand{\bea}{\begin{eqnarray}}
\newcommand{\eea}{\end{eqnarray}}
\newcommand{\nn}{\nonumber\\}

\newcommand{\eq}[1]{Eq.~\eqref{#1}}

\newcommand{\Br}{\text{Br}}
\newcommand{\BKs}{$B\to K^*\mu^+\mu^-$\ }

\begin{document}
\preprint{\vbox{\hbox{CERN-PH-TH-2015-134}\hbox{STUPP-15-222}}}

\title{Effective field theory approach to $b\to s\ell\ell^{(\prime)}$, $B\to K^{(*)}\nu\bar{\nu}$ and $B\to D^{(*)}\tau\nu$\\ with third generation couplings}

\author{Lorenzo Calibbi}
\affiliation{State Key Laboratory of Theoretical Physics, Institute of Theoretical Physics, Chinese Academy of Sciences, Beijing 100190, P.~R.~China}
\affiliation{Service de Physique Th\'eorique, Universit\'e Libre de Bruxelles,
C.P. 225, B-1050 Brussels, Belgium}
\author{Andreas Crivellin}
\affiliation{CERN Theory Division, CH--1211 Geneva 23, Switzerland}
\author{Toshihiko Ota}
\affiliation{Department of Physics, Saitama University, Shimo-Okubo 255, 338-8570 Saitama-Sakura, Japan}%

\begin{abstract}
LHCb reported anomalies in $B\to K^* \mu^+\mu^-$, $B_s\to\phi\mu^+\mu^-$ and $R(K)=B\to K \mu^+\mu^-/B\to K e^+e^-$. Furthermore, BaBar, BELLE and LHCb found hints for the violation of lepton flavour universality violation in $R(D^{(*)})=B\to D^{(*)}\tau\nu/B\to D^{(*)}\ell\nu$. In this note we reexamine these decays and their correlations to $B\to K^{(*)}\nu\bar{\nu}$ using gauge invariant dim-6 operators. For the numerical analysis we focus on scenarios in which new physics couples, in the interaction eigenbasis, to third generation quarks and lepton only. We conclude that such a setup can explain the $b\to s\mu^+\mu^-$ data simultaneously with $R(D^{(*)})$ for small mixing angles in the lepton sector (of the order of $\pi/16$) and very small mixing angles in the quark sector (smaller than $V_{cb}$). In these region of parameter space $B\to K^{(*)}\tau\mu$ and $B_s\to \tau\mu$ can be order $10^{-6}$. Possible UV completions are briefly discussed.
\end{abstract}

\maketitle

\section{Introduction}
\label{intro}

So far, the LHC completed the standard model (SM) of particle physics by discovering the last missing piece, the Higgs particle~\cite{Aad:2012tfa,Chatrchyan:2012ufa}.\footnote{We denote the SM scalar particle predicted by Brout, Englert and Higgs as the ''Higgs particle".} Furthermore, no significant direct evidence for physics beyond the SM has been found, i.e.~no new particles were discovered. However, LHCb observed indirect `hints' for new physics (NP) in $B\to K^*\mu^+\mu^-$, $B_s\to\phi\mu^+\mu^-$ and $R(K)\equiv{\rm Br}(B\to K \mu^+\mu^-)/{\rm Br}(B\to K e^+e^-)$. Furthermore, BaBar and also very recently BELLE and LHCb reported lepton flavour universality violation in $B\to D^{(*)}\tau\nu$. These observations can be used as a guideline in the exploration of possible physics beyond the SM.

In more detail, the current experimental situation is as follows: LHCb reported deviations from the SM predictions~\cite{Egede:2008uy} in $B\to K^* \mu^+\mu^-$~\cite{Aaij:2013qta,LHCb:2015dla} (mainly in an angular observable called $P_5^\prime$~\cite{Descotes-Genon:2013vna}) with a significance of $2$--$3\,\sigma$ depending on the assumptions of hadronic uncertainties~\cite{Descotes-Genon:2014uoa,Altmannshofer:2014rta,Jager:2014rwa}. Also in the decay $B_s\to\phi\mu^+\mu^-$~\cite{Aaij:2013aln} LHCb uncovered differences compared to the SM prediction based on lattice QCD~\cite{Horgan:2013pva,Horgan:2015vla} and light-cone sumrules~\cite{Straub:2015ica} of $3.1\,\sigma$ \cite{Altmannshofer:2014rta}.\footnote{Very recently, this discrepancy increased to $3.5\,\sigma$~\cite{BsphimumuFPCP}.}
Furthermore, LHCb~\cite{Aaij:2014ora} found indications for the violation of lepton flavour universality, namely
\begin{align}
	R(K)=0.745^{+0.090}_{-0.074}\pm 0.036\,,
\end{align}
in the range $1\,{\rm GeV^2}<q^2<6\,{\rm GeV^2}$. This measurement is in tension with the theoretically clean SM prediction $R_{\rm SM}(K)=1.0003 \pm 0.0001$~\cite{Bobeth:2007dw} by $2.6\,\sigma$. Combining these anomalies with all other observables for $b\to s \mu^+\mu^-$ transitions, it is found that a scenario with NP in $C_9^{\mu\mu}$ (corresponding to the operator $\bar s \gamma^\nu P_L b\,\bar\mu \gamma_\nu \mu$) but not in $C_9^{ee}$ is preferred compared to the SM by $4.3\,\sigma$~\cite{Altmannshofer:2015sma}. 

Hints for lepton flavour universality violating NP also comes from the BaBar collaboration that performed an analysis of the semileptonic $B$ decays $B\to D^{(*)}\tau\nu$~\cite{Lees:2012xj}. Recently, these decays have also been reanalyzed by BELLE~\cite{BDstaumuBELLE} and LHCb measured $B\to D^{*}\tau\nu$~\cite{BDstaumuLHCb}. In summary, these experiments have found for the ratios ${R}(D^{(*)})\equiv{\rm Br}(B\to D^{(*)} \tau \nu)/{\rm Br}(B\to D^{(*)} \ell \nu)$:
\begin{align}
R(D)_{\rm BaBar}\,&=\,0.440\pm0.058\pm0.042  \,,\\ 
R(D)_{\rm BELLE}\,&=\,0.375^{+0.064}_{-0.063}\pm0.026  \,,\\ 
R(D^*)_{\rm BaBar}\,&=\,0.332\pm0.024\pm0.018\,,\\
R(D^*)_{\rm BELLE}\,&=\,0.293^{+0.039}_{-0.037}\pm0.015  \,,\\
R(D^*)_{\rm LHCb}\,&=\,0.336\pm0.027\pm0.030  \,. 
\end{align}
Here the first (second) errors are statistical (systematic). Combining these measurements one finds~\cite{BDstaumuCOMBINED}
\begin{align}
R(D)_{\rm EXP}\,&=\,0.388\pm0.047\,,\nonumber\\ 
R(D^*)_{\rm EXP}\,&=\,0.321\pm0.021  \,. 
\label{RDEXP}
\end{align}
Comparing these measurements to the SM predictions~\cite{Fajfer:2012vx}
\begin{align}
R_{\rm SM}(D)\,&=\,0.297\pm0.017 \,,  \nonumber
\\ R_{\rm SM}(D^*) \,&=\,0.252\pm0.003 \,,
\label{RDSM}
\end{align}
we see that there is a discrepancy of 1.8\,$\sigma$ for $R(D)$ and 3.3\,$\sigma$ for $R(D^*)$ and the combination corresponds approximately to a $3.8\, \sigma$ deviation from the SM (compared to $3.4\, \sigma$ taking into account the BaBar results only~\cite{Lees:2012xj}).

Numerous models have been proposed in order to explain the anomalies in $b\to s\mu^+\mu^-$ transitions (see for example Refs.~\cite{Descotes-Genon:2013wba,Gauld:2013qba,Buras:2013qja,Gauld:2013qja,Buras:2013dea,Altmannshofer:2014cfa,Crivellin:2015mga,Crivellin:2015lwa,Niehoff:2015bfa,Sierra:2015fma,Celis:2015ara} for $Z^\prime$ models and Refs.~\cite{Becirevic:2015asa,Varzielas:2015iva} for models with leptoquarks) and the deviations from the SM predicitons in tauonic $B$ decays \cite{Crivellin:2012ye,Datta:2012qk,Crivellin:2013wna,Li:2013vlx,Faisel:2013nra,Atoui:2013zza,Sakaki:2013bfa,Celis:2013bya,Biancofiore:2014wpa}. 

Alternatively, a model independent approach using higher dimensional operators has been employed, as in the model independent fits~\cite{Descotes-Genon:2013vna,Hurth:2014vma,Altmannshofer:2014rta}. In this context, it has been argued that as $R(K)$ violates lepton flavour universality (LFU) also lepton flavour could be violated in $B$ decays~\cite{Glashow:2014iga} which might be linked to neutrino oscillations \cite{Boucenna:2015raa}.\footnote{Lepton flavour violating $B$ decays in leptoquark models have been studied in~\cite{Varzielas:2015iva} and in $Z^\prime$ models in~\cite{Crivellin:2015era}.}
While~\cite{Glashow:2014iga} considered the effect of operators at the $B$~meson scale which are invariant under electromagnetic gauge interactions only, also operators invariant under the full SM gauge group~\cite{Buchmuller:1985jz,Grzadkowski:2010es} have been considered in Ref.~\cite{Fajfer:2012jt,Bhattacharya:2014wla,Buras:2014fpa,Alonso:2015sja}.\footnote{For an analogous analysis in the lepton sector see~\cite{Crivellin:2013hpa,Crivellin:2014cta,Pruna:2014asa}.}
Here it has been claimed than an simultaneous explanation of $R(K)$, $R(D)$ and $R(D^*)$ using gauge invariant operators with left-handed fermions is possible~\cite{Bhattacharya:2014wla,Alonso:2015sja}. For this purpose, it was assumed that in the interaction eigenbasis only couplings to the third generation exist~\cite{Glashow:2014iga,Bhattacharya:2014wla} (or are enhanced by $m_\tau^2/m_\mu^2$ compared to the second one~\cite{Alonso:2015sja}), while all other couplings are generated by the misalignment between the mass and the interaction basis (or are suppressed by small lepton mass ratios~\cite{Alonso:2015sja}). 

In this article we reconsider the possibility of explaining $B\to D^{(*)}\tau\nu$ and the $b\to s\mu^+\mu^-$ data with higher dimensional gauge invariant operators, taking into account the constraints from $B\to K^{(*)}\nu\bar{\nu}$ and using the results of the global fit to $b\to s\mu\mu$ transitions. We extend the analysis of Ref.~\cite{Alonso:2015sja} and consider the possibility of lepton flavour violation (LFV) and compared to Ref.~\cite{Glashow:2014iga} we include the correlations due to $SU(2)_L$ gauge invariance and give quantitative predictions for $B\to K^{(*)}\tau\mu$ and $B_s\to \tau\mu$. 

The outline is as follows: In the next section we collect the necessary formulae for the flavour observables. Sec.~\ref{sec:GIO} discusses the gauge invariant higher dimensional operators relevant for our analysis and Sec.~\ref{sec:NA} presents our numerical results. 
Sec.~\ref{uvcomplete} briefly reviews some possible UV completions. Finally we conclude in Sec.~\ref{conclusion}.

\section{Flavour observables}
\label{flavour}
\subsection{$b\to s\mu^+\mu^-$ transitions}
$b\to s\ell_i\ell_j$ transitions are defined via the effective Hamiltonian
\bea
H_{\rm eff}^{\ell_i\ell_j} &=& - \dfrac{ 4 G_F }{\sqrt 2}
V_{tb}V_{ts}^{*} \sum\limits_{a = 9,10}
\left( {C_a^{\ell_i\ell_j} O_a^{\ell_i\ell_j} + C_a^{\prime\,\ell_i\ell_j}  O_a^{\prime\,\ell_i\ell_j} } \right)\,,\nn {O_{9(10)}^{\ell_i\ell_j}} &=& 
\dfrac{\alpha }{4\pi}[\bar s{\gamma ^\mu } P_L b]\,[\bar\ell_i{\gamma _\mu }(\gamma^5)\ell_j] \,,
\label{eq:effHam}
\eea
where the primed operators are obtained by exchanging $L\leftrightarrow R$.

Concerning \BKs, $B_s\to \phi\mu^+\mu^-$ and $B\to K\mu^+\mu^-/B\to K e^+ e^-$, as already noted in Ref.~\cite{Descotes-Genon:2013wba,Descotes-Genon:2013zva}, $C^{\mu\mu}_9<0$ and $C^{\prime\mu\mu}_9=0$ is preferred by data. However, also the possibility $C^{\mu\mu}_9=-C^{\mu\mu}_{10}<0$ gives a good fit to data. Using the global fit of Ref.~\cite{Altmannshofer:2014rta,Altmannshofer:2015sma} we see that at ($1\,\sigma$) $2\,\sigma$ level 
\bea
-0.53(-0.81) \geq & C_9^{\mu\mu}& \geq(-1.32)-1.54\,,\label{scenario1}\\
-0.18(-0.35) \geq & C_{9}^{\mu\mu}=-C_{10}^{\mu\mu}& \geq(-0.71)
-0.91\,.\label{scenario2}
\label{eq:C9fit}
\eea
Interestingly, the values of $C_9^{\mu\mu},\,C_{10}^{\mu\mu}$ favoured by $R(K)$ and \BKs lie approximately in the same range.\footnote{Note that the fit to \eq{scenario2} includes muon data only. However, as $B\to Ke^+e^-$ agrees rather well with the SM prediction, the effect on the global fit is expected to be small. Also the latest LHCb result for $B_s\to\phi\mu^+\mu^-$ \cite{BsphimumuFPCP}, which would slight increase the tension with the SM, is not included in the fit.}
Furthermore, a good fit to the current data does not require $C^{\prime\mu\mu}_9$, hence in the following we neglect operators with right-handed quark currents for simplicity. 

\subsection{$B\to K^{(*)}\nu\bar{\nu}$}

Following Ref.~\cite{Buras:2014fpa} we write the relevant effective Hamiltonian as
\begin{align}
{H_{\rm eff}^{\nu_i\nu_j} } &=  - \frac{{4{G_F}}}{{\sqrt 2 }}{V_{tb}}V_{ts}^*\left( {{C_L^{ij}}{O_L^{ij}} + {C_R^{ij}}{O_R^{ij}}} \right)\,\\
O_{L,R}^{ij} &= \frac{\alpha }{{4\pi }} [\bar s{\gamma ^\mu }{P_{L,R}}b][{{\bar \nu }_i}{\gamma _\mu }\left( {1 - {\gamma ^5}} \right){\nu _j}]\,,
\end{align}
and $C_L^{\rm SM}\approx-1.47/s_w^2$. In the limit of vanishing right-handed $sb$ current, the branching ratios normalized by the SM predictions read
\begin{equation}
{R_{K^{(*)}}^{\nu\bar{\nu}}} = 
\frac{1}{3}\sum\limits_{i,j=1}^3 \dfrac{ \left|  {C_L^{ij}} \right|^2}{\left| {C_L^{\rm SM}} \right|^2} \,.
\end{equation}
The current experimental limits are ${R_K^{\nu\bar{\nu}}} < 4.3$~\cite{Lees:2013kla} and ${R_{{K^*}}^{\nu\bar{\nu}}} < 4.4$~\cite{Lutz:2013ftz}.

\subsection{$B\to D^{(*)}\tau\nu$}

The effective Hamiltonian for semileptonic $b\to c$ transitions is
\begin{equation}
{H_{\text{eff}}} = \frac{{4{G_F}}}{{\sqrt 2 }}{V_{cb}}C_{L\,ij}^{cb} [\bar c{\gamma ^\mu }{P_L}b][\bar \ell_i {\gamma _\mu }P_L\nu_j]\,,
\end{equation}
with $C_{L\,ij}^{cb\,{\rm SM}}=\delta_{ij}$ (for massless neutrinos) taking into account only left handed vector currents. In this case the ratios of branching ratios are 
\begin{equation}
\dfrac{R(D^{(*)})_{\rm EXP}}{R(D^{(*)})_{\rm SM}} = 
 \dfrac{{\sum\limits_{j=1}^3} {{{\left| {C_{L\;3j}^{cb}} \right|}^2}}}{{\sum\limits_{j=1}^3}{{{\left| {C_{L\;\ell j}^{cb}} \right|}^2}}}\,,
\end{equation}
with $\ell=e,\mu$ which has to be compared to \eq{RDSM} and \eq{RDEXP}.

\subsection{Lepton-flavour violating \boldmath$B$\unboldmath\ decays}
\label{susec:H}

Here we give formulas for the branching ratios of LFV $B$ decays following the analysis of Ref.~\cite{Crivellin:2015era}. We take into account only contributions from the operators $O_{9}^{(\prime)\ell\ell^\prime}$ and $O_{10}^{(\prime)\ell\ell^\prime}$ while neglecting contributions from operators with scalar currents not relevant for our analysis. For $B_s\to\ell^+\ell^{\prime -}$ (with $\ell\neq \ell^\prime$) we use the results of Ref.~\cite{Dedes:2008iw} neglecting the mass of the lighter lepton. The branching ratios for $B\to K^{(*)}\tau^\pm\mu^\mp, B\to K^{(*)} \mu^\pm e^\mp$ are computed using form-factors obtained from lattice QCD in Ref.~\cite{Bouchard:2013eph} (see also Refs.~\cite{Horgan:2015vla,Horgan:2013hoa}). The final results read
\begin{widetext}
\bea
\Br\left[ B_s \to \ell^+ \ell^{\prime-} \right] &=& \dfrac{\tau_{B_s}
  m_\ell^2 M_{B_s} f_{B_s}^2}{32\pi^3} \alpha^2 G_F^2 \left| V_{tb}
V_{ts}^*\right|^2 \left(1 - \dfrac{{\rm
    Max}[m_\ell^2,m_{\ell^\prime}^2]}{M_{B_s}^2}\right)^2 \left(\left|
C_{9}^{\ell \ell^\prime}-C^{\prime\ell \ell^\prime}_{9}\right|^2 +
\left| C_{10}^{\ell \ell^\prime}-C^{\prime\ell \ell^\prime}_{10}
\right|^2 \right) ,\nonumber\\
\Br[B\to K^{(*)}\ell^+\ell^{\prime-}] &=& 10^{-9} \left(a_{K^{(*)}\ell\ell^\prime}\left|C_9^{\ell\ell^\prime} + C_9^{\prime\ell\ell^\prime} \right|^2 +
b_{K^{(*)}\ell\ell^\prime}\left|C_{10}^{\ell\ell^\prime} + C_{10}^{\prime\ell\ell^\prime} \right|^2 \right. \nn &\;\;\;&+\left.
c_{K^*\ell\ell^\prime}\left|C_9^{\ell\ell^\prime}
-C_9^{\prime\ell\ell^\prime} \right|^2 +d_{K^*\ell\ell^\prime}\left|C_{10}^{\ell\ell^\prime}-C_{10}^{\prime\ell\ell^\prime} \right|^2 \right)\,,
\label{bkstaumu}
\eea
with
\begin{center}
\begin{tabular}{|c|c|c|c|c|c|c|}
\hline
$\ell\ell^\prime $ & $a_{K\ell\ell^\prime}$ & $b_{K\ell\ell^\prime}$ &
$a_{K^*\ell\ell^\prime}$ & $b_{K^*\ell\ell^\prime}$ &
$c_{K^*\ell\ell^\prime}$ & $d_{K^*\ell\ell^\prime}$ \\
\hline
$\;\tau\mu,\tau e\;$ & $\;9.6 \pm 1.0\;$ & $\;10.0 \pm 1.3\;$ & $\;3.0 \pm
0.8\;$ & $\;2.7 \pm 0.7\;$ & $\;16.4 \pm 2.1\;$ & $\;15.4 \pm 1.9\;$
\\
$\mu e$ & $15.4 \pm 3.1$ & $15.7 \pm 3.1$ & $5.6 \pm 1.9$ & $5.6 \pm
1.9$ & $29.1 \pm 4.9$ & $29.1 \pm 4.9$ \\
\hline
\end{tabular}
\end{center}
\end{widetext}
The formula for the branching ratio of $B_s \to \ell^+ \ell^{\prime-}$ is symmetric with respect to the exchange of 
$C_9^{(\prime)\ell\ell^\prime}\leftrightarrow C_{10}^{(\prime)\ell\ell^\prime}$, while in the case of $B \to K^{(*)}\ell^+ \ell^{\prime-}$
this symmetry is broken by lepton-mass effects. There is a small difference between the theoretical prediction for 
the charged mode $B^+\to K^{(*)+}\ell^+ \ell^{\prime-}$ and the neutral one $B^0\to K^{(*)0}\ell^+ \ell^{\prime-}$ due to the different $B$-meson lifetime $\tau_B$ which we neglected fixing the numerical value of $\tau_B$ to the one of the neutral meson. Note that the results above are given for $\ell^-\ell^{\prime+}$ final states and not for the sum $\ell^\pm\ell^{\prime\mp}=\ell^-\ell^{\prime+}+\ell^+\ell^{\prime-}$ to which the experimental constraints apply~\cite{Amhis:2014hma}. The only channel with $\tau\mu$ final states for which an experimental upper limit exists is
\bea
{\rm Br}\left[B^+\to K^+\tau^\pm\mu^\mp \right]_{\rm exp}\le 4.8\times 10^{-5} \,.
\eea

\section{Gauge invariant operators}
\label{sec:GIO}

As we have previously seen, a scenario with left-handed currents only gives a good fit to data, cf.~\eq{scenario2}. In such a scenario $SU(2)_L$ relations are necessarily present. These relations are automatically taken into account once gauge invariant operators are considered. Therefore, let us focus on 4-fermion operators with left-handed quarks and leptons. There are two such 4-fermion operators in the effective Lagrangian
\begin{equation}
	{\cal L_{\rm dim6}}=\dfrac{1}{\Lambda^2}\sum{O_X C_X}\,,
\end{equation}
where $\Lambda$ is the scale of NP, which can contribute to $b\to s\ell\ell$ transitions at tree-level \cite{Buchmuller:1985jz,Grzadkowski:2010es}:
\begin{align}
\label{Q1-Q3}
Q_{\ell q}^{\left( 1 \right)} = \left( {\bar L{\gamma ^\mu }L} \right)\left( {\bar Q{\gamma _\mu }Q} \right)\,,~~
Q_{\ell q}^{\left( 3 \right)} = \left( {\bar L{\gamma ^\mu }{\tau _I}L} \right)\left( {\bar Q{\gamma _\mu }{\tau ^I}Q} \right)\,,
\end{align}
where $L$ is the lepton doublet and $Q$ the quark doublet and the flavour indices are not explicitly shown here. Writing these operators in terms of their $SU(2)_L$ components (i.e.~up-quarks, down-quarks, charged leptons and neutrinos) we find for the terms relevant for the processes discussed in the last section (before EW symmetry breaking)
\begin{align}
{\cal L} \supset & \frac{{C_{ijkl}^{\left( 1 \right)}}}{{{\Lambda ^2}}}\left( {{{\bar \ell }_i}{\gamma ^\mu }{P_L}{\ell _j}{{\bar d}_k}{\gamma _\mu }{P_L}{d_l} + {{\bar \nu }_i}{\gamma ^\mu }{P_L}{\nu _j}{{\bar d}_k}{\gamma _\mu }{P_L}{d_l}} \right) +\nonumber\\
 &\frac{C_{ijkl}^{\left( 3 \right)}}{{{\Lambda ^2}}}\left( {2{{\bar \ell }_i}{\gamma ^\mu }{P_L}{\nu _j}{{\bar u}_k}{\gamma _\mu }{P_L}{d_l} - {{\bar \nu }_i}{\gamma ^\mu }{P_L}{\nu _j}{{\bar d}_k}{\gamma _\mu }{P_L}{d_l}}\right.\nonumber\\ 
 &+\left.{ {{\bar \ell }_i}{\gamma ^\mu }{P_L}{\ell _j}{{\bar d}_k}{\gamma _\mu }{P_L}{d_l}} \right)\,,
\end{align}
where $C^{(1,3)}_{ijkl}$ are the dimensionless coefficients of the operators of Eq.~(\ref{Q1-Q3}). After EW symmetry breaking the following redefinitions of the fields are performed in order to render the mass matrices diagonal
\begin{equation}
{d_L} \to D_{}^\dag {d_L},\;{u_L} \to U_{}^\dag {u_L},\;\;{\ell _L} \to L_{}^\dag {\ell _L},\;\nu  \to L_{}^\dag \nu \,.
\end{equation}
We define for future convenience
\begin{equation}
\lambda^{(1,3)}\tilde X_{ij}^{\left( {1,3} \right)}\tilde Y_{kl}^{\left( {1,3} \right)} = L_{i'i}^*{L_{j'j}}D_{k'k}^*{D_{l'l}}C_{i'j'k'l'}^{\left( {1,3} \right)}\,,
\end{equation}
where $\lambda^{(1,3)}$ are overall constants. Using constraints from the measured CKM matrix, i.e.~$V = U^\dag D$, we finally obtain
\begin{align}
C_9^{ij} =&  -C_{10}^{ij}
\nonumber 
\\ =& \frac{\pi }{{\sqrt 2 {\Lambda ^2}{G_F}\alpha {V_{tb}}V_{ts}^*}}\left( {{\lambda ^{(1)}}\tilde X_{ij}^{(1)}\tilde Y_{23}^{(1)} + {\lambda ^{(3)}}\tilde X_{ij}^{(3)}\tilde Y_{23}^{(3)}} \right)\nonumber\\
C_L^{ij} 
=& 
\frac{\pi }{{\sqrt 2 {\Lambda ^2}{G_F}\alpha {V_{tb}}V_{ts}^*}}\left( {{\lambda ^{(1)}}\tilde X_{ij}^{(1)}\tilde Y_{23}^{(1)} - {\lambda ^{(3)}}\tilde X_{ij}^{(3)}\tilde Y_{23}^{(3)}} \right)\nonumber\,,
\\
C_{L\;ij}^{cb} &=  - {\frac{\lambda^{(3)}}{{\sqrt 2 {\Lambda ^2}{G_F}}}}\frac{{{{\tilde X}_{ij}^{(3)}}}}{{{V_{cb}}}}\sum\limits_k {\left( {{V_{2k}}{{\tilde Y}_{k3}^{(3)}}} \right)}\,,\label{C9C10CL}
\end{align}
for the Wilson coefficients relevant for $b\to s\mu^+\mu^-$,  $B\to K^{(*)}\nu\bar{\nu}$ and $B\to D^{(*)}\tau\nu$ respectively.
Note that in the limit $C^{(1)}=C^{(3)}$ the contribution to $B\to K^{(*)}\nu\bar{\nu}$ vanishes. 

\begin{figure}[t]
\begin{center}
\begin{tabular}{cp{7mm}c}
\includegraphics[width=0.43\textwidth]{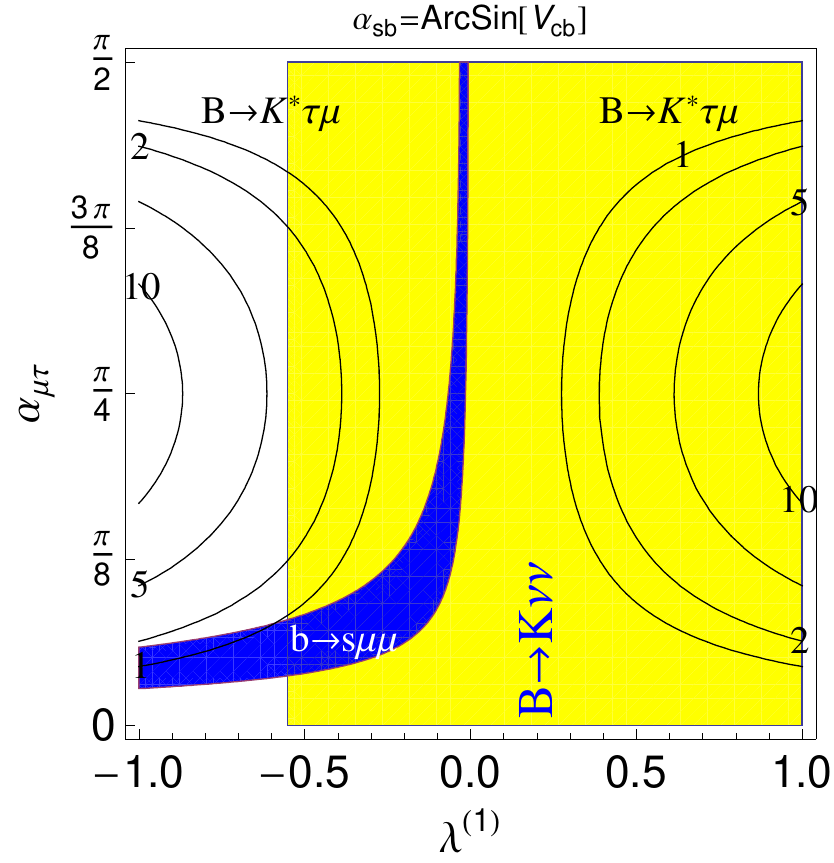} &&
\end{tabular}
\end{center}
\caption{Allowed regions in the $\lambda^{(1)}$--$\alpha_{\mu\tau}$ plane from
 $b\to s \mu^+\mu^-$ data (blue) and $B\to K \nu\bar{\nu}$ (yellow) for $\alpha_{sb}={\rm ArcSin}[V_{cb}]$ and $\Lambda=1\,$TeV. Note that here changing $\alpha_{sb}$ only has the effect of an overall scaling of $\lambda^{(1)}$. The contour lines denote ${\rm Br}[B\to K^*\tau\mu]$ in units of $10^{-6}$.
\label{Plot1}}
\end{figure}

\section{Numerical analysis}
\label{sec:NA}
\begin{figure*}[t]
\begin{center}
\begin{tabular}{cp{7mm}c}
\includegraphics[width=0.32\textwidth]{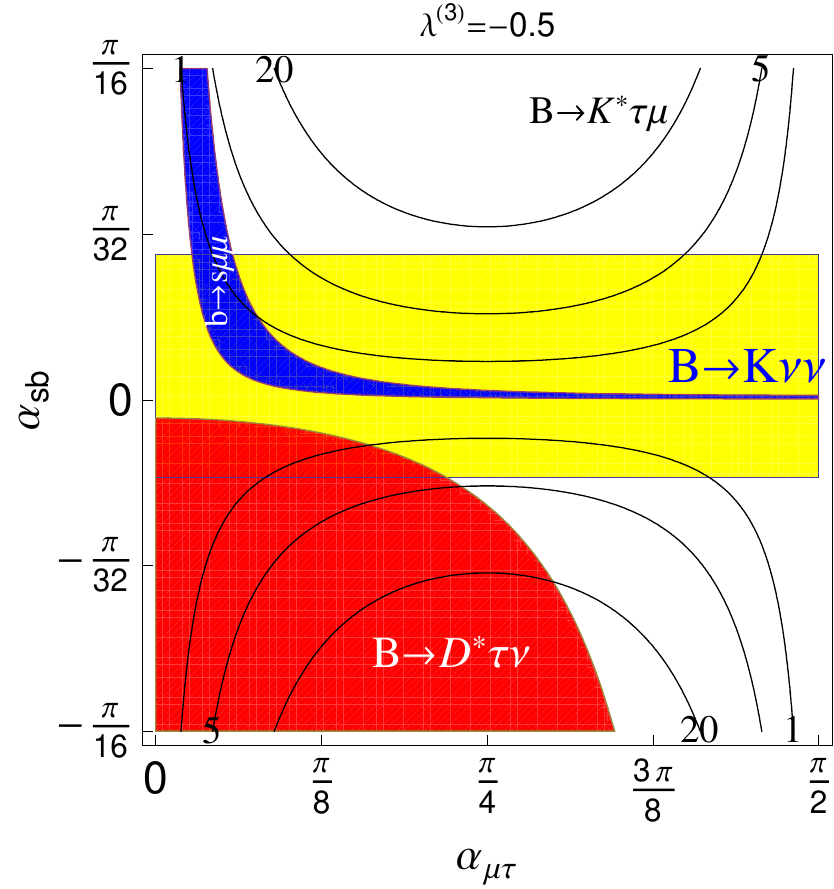}
\includegraphics[width=0.32\textwidth]{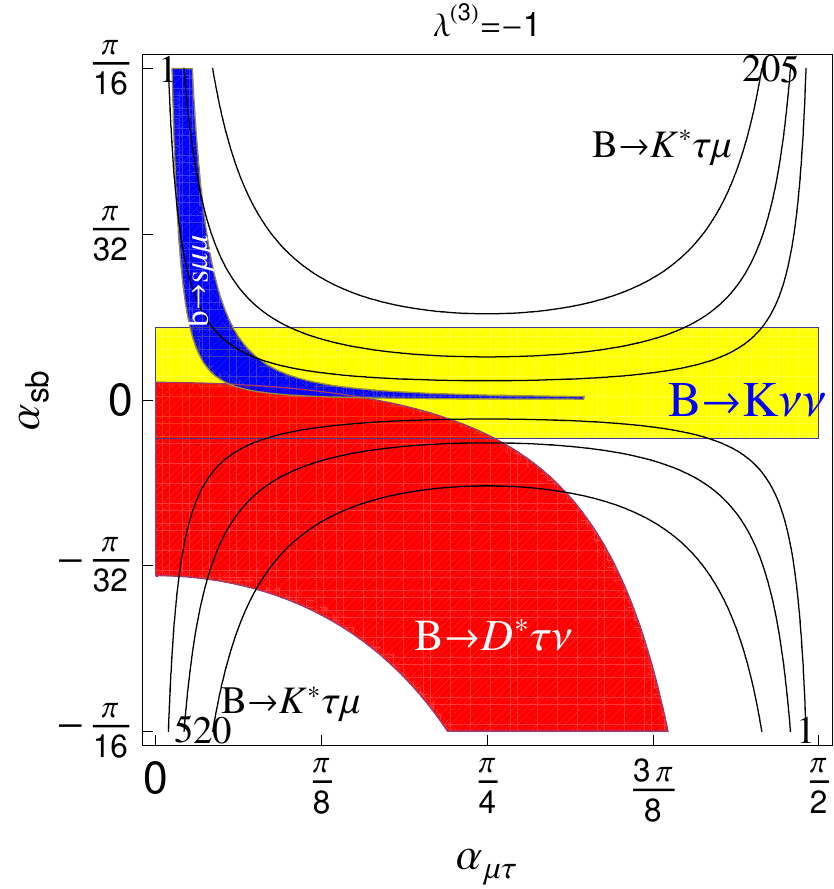}
\includegraphics[width=0.32\textwidth]{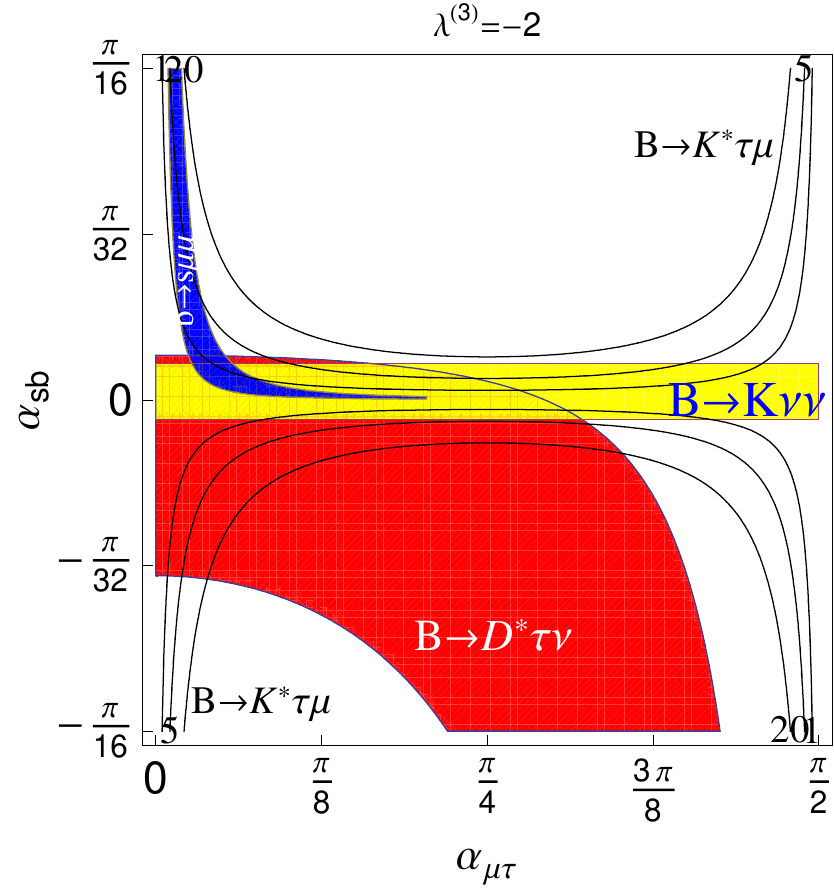}
\end{tabular}
\end{center}
\caption{Allowed regions in the $\alpha_{\mu\tau}$--$\alpha_{sb}$ plane from $B\to K\nu\bar{\nu}$ (yellow), $R(D^*)$ (red) and $b\to s \mu^+\mu^-$ (blue) for $\Lambda=1\,$TeV and $\lambda^{(3)}=-0.5$ (left plot), $\lambda^{(3)}=-1$ (middle) and $\lambda^{(3)}=-2$ (right). Note that $\alpha_{sb}=\pi/64$ roughly corresponds to the angle needed to generate $V_{cb}$ and that if $\lambda^{(3)}$ is positive, $R(D^*)$ and $b\to s\mu^+\mu^-$ cannot be explained simultaneously. \label{Plot2}}
\end{figure*}
Since we have $C_9^{\tau\mu}=-C_{10}^{\tau\mu}$ we find for the LFV $B$ decays
\begin{align}
{\rm Br}\left[B\to K\tau^\pm\mu^\mp \right]/{\rm Br}\left[B\to K^*\tau^\pm\mu^\mp \right]&\approx 1\,,\\
{\rm Br}\left[B_s\to \tau^\pm\mu^\mp \right]/{\rm Br}\left[B\to K^*\tau^\pm\mu^\mp \right]&\approx 0.5\,.
\end{align}
Therefore in the following, we will just present the numerical evaluation of ${\rm Br}\left[B\to K^*\tau^\pm\mu^\mp \right]$ while ${\rm Br}\left[B_s\to \tau^\pm\mu^\mp \right]$ and ${\rm Br}\left[B\to K\tau^\pm\mu^\mp \right]$ can be obtained by the appropriate rescaling.

We also note that $B\to K\nu\bar{\nu}$ imposes an upper limit on the absolute value of $C_9^{\tau\tau}=-C_{10}^{\tau\tau}$ and $C_9^{\tau\mu}=-C_{10}^{\tau\mu}$ valid for $C^{(3)}$ and $C^{(1)}$ separately. Neglecting the small NP contribution to $C_L^{\mu\mu}$ and assuming no NP in the electron channel we find:
\begin{align}
\dfrac{|C_9^{\tau\mu}|}{C_L^{\rm SM}}&\le\sqrt{4.3\times3/2}\approx 2.5\,,\label{C9tautau}\\
\dfrac{|C_9^{\tau\tau}|}{C_L^{\rm SM}}&\le\sqrt{3\times4.3\times3/2-2}+1\approx 5.2\,.
\end{align}
This leads to the following upper limits valid in any model generating only $C^{(3)}$ or $C^{(1)}$:
\begin{align}
{\rm Br}[B\to K\tau\mu]\leq 8.3\times 10^{-6}\,.
\end{align}
However, this limit can be evaded for $C^{(3)}=C^{(1)}$. In Ref.~\cite{Alonso:2015sja} it was proposed that the MFV-like relation $\tilde Y_{22}/\tilde Y_{33}=m_\tau^2/m_\mu^2$ could explain $R(D^{(*)})$ and $b\to s\mu^+\mu^-$ data simultaneously. From \eq{C9tautau} we see that this ansatz is only possible for $C^{(3)}=C^{(1)}$ but not if $C^{(3)}$ or $C^{(1)}$ are separately different from zero.

Therefore, we will focus in the following on scenarios with third generation couplings in the EW basis only, which correspond to a general rank 1 matrix in the mass eigenbasis, as suggested in Ref.~\cite{Glashow:2014iga,Bhattacharya:2014wla}. In other words we have
\begin{align}
C_{ijkl}^{(1,3)}&=\lambda^{(1,3)}\tilde X_{ij}\tilde Y_{kl}\,,\\
\tilde X &= {L^\dag }XL,\;\tilde Y = {D^\dag }YD\,,\;\;\;
X &= Y = \left( {\begin{array}{*{20}{c}}0&0&0\\0&0&0\\0&0&1\end{array}} \right)\,.\nonumber
\end{align}
Taking into account only rotations among the second and third generation one finds
\begin{align}
\tilde X =& \left( {\begin{array}{*{20}{c}}
0&0&0\\
0&{{{\sin }^2}\left( \alpha_{\mu\tau}  \right)}&{ - \sin \left( \alpha_{\mu\tau}  \right)\cos \left( \alpha_{\mu\tau}  \right)}\\
0&{ - \sin \left( \alpha_{\mu\tau}  \right)\cos \left( \alpha_{\mu\tau}  \right)}&{{{\cos }^2}\left( \alpha_{\mu\tau}  \right)}
\end{array}} \right)\,,\nonumber\\
\tilde Y =& \left( {\begin{array}{*{20}{c}}
0&0&0\\
0&{{{\sin }^2}\left( \alpha_{sb}  \right)}&{ - \sin \left( \alpha_{sb}  \right)\cos \left( \alpha_{sb}  \right)}\\
0&{ - \sin \left( \alpha_{sb}  \right)\cos \left( \alpha_{sb}  \right)}&{{{\cos }^2}\left( \alpha_{sb}  \right)}
\end{array}} \right)\,.\nonumber
\end{align}
Note that a rotation $\sin(\alpha_{sb})\gg V_{cb}$ would require fine-tuning with the up sector in order to obtain the correct CKM matrix.

\subsubsection{$Q_{\ell q}^{\left( 1 \right)}$ operator}
In this case we have neutral currents only. As a consequence, there is obviously no effect in $R(D^{(*)})$, but $b\to s \mu^+\mu^-$ is directly correlated to $B\to K^{(*)}\nu\bar{\nu}$ depending on the angle $\alpha_{\mu\tau}$. Note that a change in $\alpha_{sb}$ can be compensated by a change in $\lambda^{(1)}$ and therefore does not affect the correlations among $B\to K^{(*)}\nu\bar{\nu}$ and $b\to s \mu^+\mu^-$ transitions. In Fig.~\ref{Plot1} the regions favoured by $b\to s\mu^+\mu^-$ (blue) and allowed by $B\to K\nu\bar{\nu}$ (yellow) are shown together with contour lines for $B\to K^*\tau\mu$ in units of $10^{-6}$. Note that $B\to K\nu\bar{\nu}$ rules out branching ratios for $B\to K^*\tau\mu$ above approximately $1\times10^{-6}$ and that the constraint from $B\to K\nu\bar{\nu}$, being inclusive in the neutrino flavours, is independent of $\alpha_{\mu\tau}$.

\subsubsection{$Q_{\ell q}^{\left( 3 \right)}$ operator}
Here we have also charged currents that are related to the neutral current processes via CKM rotations. In Fig.~\ref{Plot2} 
the regions allowed by $B\to K\nu\bar{\nu}$ (yellow) and giving a good fit to data for $b\to s\mu^+\mu^-$ (blue) and (at the $2\,\sigma$ level) for $B\to D^*\tau\nu$ (red) are shown for different values of $\lambda^{(3)}$. Note that $b\to s\mu^+\mu^-$ data can be explained simultaneously with $R(D^{(*)})$ for negative $\mathcal{O}(1)$ values of $\lambda^{(3)}$ without violating the bounds from $B\to K\nu\bar{\nu}$. Again, in the regions compatible with all experimental constraints, the branching rations of LFV $B$ decays to $\tau\mu$ final states can only be up to $\approx10^{-6}$.

\subsubsection{$Q_{\ell q}^{\left( 1 \right)}$ and $Q_{\ell q}^{\left(3 \right)}$ with $\lambda^{(1)}=\lambda^{(3)}$}
\begin{figure}[tb]
\begin{center}
\begin{tabular}{cp{7mm}c}
\includegraphics[width=0.43\textwidth]{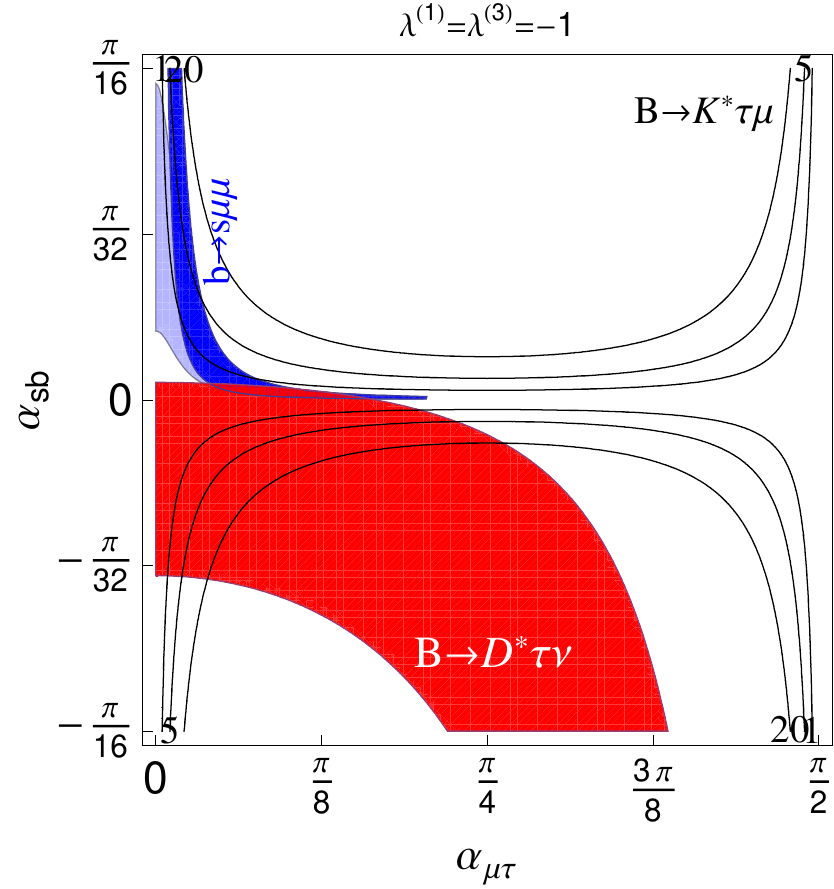} &&
\end{tabular}
\end{center}
\caption{
Allowed regions in the $\alpha_{\mu\tau}$--$\alpha_{sb}$ plane from $R(D^*)$ (red) and $b\to s \mu^+\mu^-$ (dark blue) for $\Lambda=\,$TeV and $\lambda^{(3)}=\lambda^{(1)}=-1$. The light blue region corresponds to the MFV-like ansatz for the lepton masses. Note that $\alpha_{sb}=\pi/64$ roughly corresponds to the angle needed to generate $V_{cb}$ and that the MFV-like Ansatz only differs marginally from the one with third generation couplings only in the region compatible with $R(D)$. The contour lines denote ${\rm Br}[B\to K^*\tau\mu]$ in units of $10^{-6}$.           
\label{PlotC3C1}}
\end{figure}

In this case the phenomenology is then rather similar to the case of $C^{(3)}$ only. The major differences are that, as already mentioned before, the bounds from $B\to K\nu\bar{\nu}$ are evaded and the relative contribution to $b\to s\mu\mu$ compared to $R(D^{(*)})$ is a factor of 2 larger. In Fig.~\ref{PlotC3C1} we show the analogous plot to the central panel of Fig.~\ref{Plot2} ($\lambda^{(3)}=\lambda^{(1)}=-1$) for this scenario. Note that again $R(D^{(*)})$ rules out very large branching ratios for lepton flavour violating $B$ decays in the regions compatible with $b\to s\mu^+\mu^-$ data. We also consider the MFV-like ansatz \cite{Alonso:2015sja} with additional flavour rotations (light blue) which however differs only slightly for the ansatz with third generation couplings.

\section{UV completions}
\label{uvcomplete}
Let us briefly discuss UV completions which can give the desired coupling structure. As discussed previously, the 4-Fermi operator $Q_{\ell q}^{(3)}$ is relevant both for $R(K)$ and $R(D^{(*)})$. If $Q_{\ell q}^{(3)}$ is mediated by a single field, then there are only four possibilities:
(i) Vector boson (VB) with the SM charges $(SU(3)_{c}, SU(2)_{L}, U(1)_{Y})=({\bf 1},{\bf 3},0)$,
(ii) Scalar leptoquark (SLQ) with ({\bf 3},{\bf 3},$-1/3$),
(iii) Vector leptoquark (VLQ) with ({\bf 3},{\bf 1},2/3),
and (iv) Vector leptoquark with ({\bf 3},{\bf 3},2/3). 
The vector boson ({\bf 1},{\bf 3},0) induces only $Q_{\ell q}^{(3)}$.
On the other hand, the leptoquark fields result in 
particular combinations of $Q_{\ell q}^{(1)}$ and 
$Q_{\ell q}^{(3)}$~\cite{Alonso:2015sja}.
With the assumption of the third generation coupling,
the relative size of the effective couplings $\lambda^{(1,3)}$
and the signs are determined as 
\begin{align}
\text{VB({\bf 1},{\bf 3},$0$) : }& 
\lambda^{(3)}\text{ both positive and negative} ,
\end{align}
\begin{align}
\text{SLQ({\bf 3},{\bf 3},$-1/3$) : }& 
\lambda^{(1)} = 3 \lambda^{(3)},  
 \quad 
\lambda^{(3)}>0,
\\
\text{VLQ({\bf 3},{\bf 1},2/3) : }&
\lambda^{(1)} = \lambda^{(3)},
\quad 
\lambda^{(3)} < 0,
\\
\text{VLQ({\bf 3},{\bf 3},2/3) : }&
\lambda^{(1)} = -3 \lambda^{(3)},
\quad
\lambda^{(3)}
>0.
\end{align}
The coefficient $C_{9}^{ij}$ is proportional to $\lambda^{(1)}+\lambda^{(3)}$ and a negative value is favoured by $R(K)$. Therefore, the scalar leptoquark is rejected as a candidate. To explain $R(D^{(*)})$ simultaneously, $\lambda^{(3)}$ itself must also be negative. This condition excludes the triplet vector leptoquark. If the experimental results are explained by the operator $Q_{\ell q}^{(3)}$ under the assumption of third generation coupling only, the possible mediators are the triplet vector boson or the singlet vector leptoquark.
According to the analysis of the previous section, a good fit to flavour data requires a mediator mass of $\mathcal{O}$(1) TeV. 
This opens interesting prospects for the LHC, especially in the case of leptoquarks that can be produced in proton-proton collisions 
via colour interactions and would decay to one lepton ($\tau$ or more interestingly $\mu$) and one jet (possibly a $b$-jet).

\section{Conclusions}
\label{conclusion}
In this article we considered the effect of gauge invariant dim-$6$ operators with left-handed fermions on $b\to s\mu^+\mu^-$, $B\to K^{(*)}\nu\bar{\nu}$, $B\to D^{(*)}\tau\nu$, $B\to K^{(*)}\tau\mu$ and $B_s\to \tau\mu$. For operators with left-handed quarks and leptons we find the correlations ${\rm Br}\left[B\to K\tau^\pm\mu^\mp \right]\approx {\rm Br}\left[B\to K^*\tau^\pm\mu^\mp \right]\approx 2 {\rm Br}\left[B_s\to \tau^\pm\mu^\mp \right]$. We showed that the anomalies in $b\to s\mu\mu$ data can be explained simultaneously with $R(D^*)$. For this we considered scenarios in which third generation couplings in the EW basis are present only: $\lambda^{(1)}\neq0$, $\lambda^{(3)}\neq0$ and $\lambda^{(3)}=\lambda^{(1)}\neq0$. Taking into account $\lambda^{(1)}\neq0$ only, $b\to s\mu^+\mu^-$ data can be explained without violating bounds from $B\to K^{(*)}\nu\bar{\nu}$. However, in the allowed regions of parameter space, ${\rm Br}[B\to K^{(*)}\tau\mu]$ can only be up to $1\times 10^{-6}$. In the case of $\lambda^{(3)}\neq0$, $b\to s\mu^+\mu^-$ data can be explained simultaneously with $R(D^*)$. In these regions ${\rm Br}[B\to K^{(*)}\tau\mu]$ can again be only up to $10^{-6}$. Finally we considered $\lambda^{(3)}=\lambda^{(1)}\neq0$. Such a scenario can be realized with a leptoquark in the singlet representation of $SU(2)_L$ (making an MFV-like ansatz for the lepton couplings possible) and constraints from $B\to K^{(*)}\nu\bar{\nu}$ are avoided. Again, LFV $B$ decays turn out to be of the same order as in the other scenarios.

{\it Note added} --- During the completion of this work, an article presenting a dynamical model with additional vector bosons and third generation couplings appeared in which $Q_{\ell q}^{(3)}$  is generated~\cite{Greljo:2015mma}.

{\it Acknowledgments} --- {\small A.C.~and T.O.~thank the ULB for hospitality during their visit in Brussels. A.C.~is supported by a Marie Curie Intra-European Fellowship of the European Community's 7th Framework Programme under contract number PIEF-GA-2012-326948. T.O.~is supported by Japan Society for the Promotion of Science under KAKENHI Grant Number 26105503.
L.C.~thanks the Munich Institute for Astro- and Particle Physics and the organizers of the workshop ``Indirect Searches for New Physics in the LHC and Flavour Precision Era'' for hospitality and partial financial support during the completion of this work.}

\bibliography{BIB}

\end{document}